\documentclass{PoS}
\usepackage{graphicx}

\newcommand{\Tr}{\mathop{\textrm{Tr}}}
\newcommand{\exv}[1]{\left\langle{#1}\right\rangle}
\newcommand{\x}{{\bf x}}
\renewcommand{\r}{{\bf r}}
\renewcommand{\d}{\partial}

\title{Static quark free energies at finite temperature}

\ShortTitle{Quark free energy}

\author{Z. Fodor$^{1,2}$, A. Jakov\'ac$^2$, S.D.~Katz$^{1,2}$, K.K. Szab\'o$^2$\\
$^1$Institute for Theoretical Physics, E\"otv\"os University, H-1117 Budapest, Hungary. \\
$^2$Department of Physics, University of Wuppertal, D-42097 Wuppertal, Germany.\\
       E-mail: \email{jakovac@esr.phy.bme.hu}
}

      \abstract{We determine the static quark free energies around the
        transition temperature using 2+1 flavors of staggered fermions.
        Simulations are carried out on $N_t=4,\,6,\,8$ and $10$ lattices using
        physical quark masses. The free energies extracted from Polyakov-loop
        correlators are extrapolated to the continuum limit.  }

\FullConference{The XXV International Symposium on Lattice Field Theory\\
		 July 30-4 August 2007\\
		 Regensburg, Germany}

\begin{document}

\section{Introduction}

Our aim is to compute the free energy of a static quark-antiquark pair. There
are several measurements on this quantity in the literature (for recent
publications cf.
\cite{Kaczmarek:2002mc,Bornyakov:2003ev,Kaczmarek:2004gv,Maezawa:2007fc}).
Here we go beyond these computations, we use physical quark masses and perform
a careful continuum limit extrapolation with the necessary renormalization
procedure.

The quark-antiquark free energy can be expressed 
as correlators of Polyakov loops:
\begin{equation}
  e^{-F_{\bar{q}q}(\r)/T} \sim \sum\limits_\x \exv{\Tr P(\x)\,\Tr P^\dagger(\x+\r)},
\end{equation}
where \r\ is a vector in the spacial direction, $T=1/(N_t a)$ is the
temperature and $\x$ runs over all the spatial lattice sites. $P$ is the
Polyakov loop
\begin{equation}
  P(\x) = \prod\limits_{x_4=0}^{N_t-1} U_4(\x,x_4),
\end{equation}
where $U_\mu(x)\in SU(3)$ is the link gauge field. 

In pure gauge theory we expect that the Polyakov loop correlator behaves
Coulomb-like at short distances. In the deconfined phase the Coulomb behavior
is screened at large distances, the exponential range defines the screening
mass. In the confined phase the free energy is linearly rising, the derivative
of the rise is the string tension. This behavior can give an
account for the quark confinement and Regge trajectories at zero temperature.

The above picture is modified, however, when we include dynamical quarks. At
large distances it is favorable to generate a quark-antiquark pair from the
vacuum, which then screens the color field between the two Polyakov
loops~\cite{Bali:2005fu}.  From this point (the string breaking scale) the
lowest energy level will be insensitive of the position of the heavy quarks,
resulting in a constant free energy.  The value of this constant restricts the
possible bound state energies, calculated in the given potential, as no bound
state can be formed with energy larger then the maximum energy.

At finite temperature the above picture persists, but we can also have general
expectations about the temperature dependence. Physically we expect that in a
thermal vacuum it is easier to generate a quark-antiquark pair than at $T=0$, 
since there
are thermally excited particles around which can scatter on the gluonic 
string
between the static quark-antiquark pair. The gluonic string, being excited
itself, can more easily break into a dynamical quark-antiquark pair.
This suggests that the string breaking scale and so
the flattened free energy value decreases with the temperature. This dynamical
picture coincides with the thermodynamical expectation. The negative
temperature derivative of the free energy is the entropy, which must
be positive in a stable system:
\begin{equation}
  \left. - \frac{\d F_{\bar{q}q}(\r,T)}{\d T}\right|_{V} = S_{\bar{q}q}(\r,T) > 0.
\end{equation}
This formula should be true for any \r, since \r\ here is just a parameter,
telling the position of the fixed Polyakov loops. As a consequence we expect
that at any point the quark-antiquark free energy decreases with the
temperature. This condition can be an important check for the correctness of
the renormalization procedure.

String breaking effects compete with screening. If the free energy is screened
before it can rise to the string breaking scale, then screening wins,
otherwise the string breaking effect. But the main features of the free energy
are the same in both cases. Since there is no phase transition in QCD, the two
regimes are connected with each other continuously.

\section{Renormalization}

When we approach the continuum limit, the value of the unrenormalized free
energy diverges. This is because in a single Polyakov loop the self-energy is
divergent. We expect:
\begin{equation}
  \exv{\Tr P(x)}\biggr|_\mathrm{div} = e^{-C(a) N_ta} = e^{-C(a)/T},
\end{equation}
where $C(a)\to\infty$ in the continuum limit. At finite $a$ the specific value
of $C(a)$ has no physical meaning, since it depends on how we define the
``divergent part'' of the self-energy (renormalization scheme). Although the
constant $C(a)$ can be chosen in different ways, it is important that it
should only depend on the lattice spacing. In the literature there are several
ways to fix this constant \cite{Kaczmarek:2004gv,Fodor:2005qy}.

Subtracting the divergent part from the free energy, the renormalized free
energy can be defined as
\begin{equation}
  e^{-F^\mathrm{ren}_{\bar{q}q}(\r,a)/T} =e^{-F_{\bar{q}q}(\r,a)/T} e^{2C(a)/T},\qquad
  \Rightarrow \qquad F^\mathrm{ren}_{\bar{q}q}(\r,a) = F_{\bar{q}q}(\r,a) - 2C(a).
\end{equation}

A possible way of fixing $C(a)$ is to take a physical observable based on
$F_{\bar{q}q}$, and requiring that it should be independent of $a$. We
emphasize that there is no restriction on the physical quantity other than it
must be fixed and be finite if $F_{\bar{q}q}$ is finite. It needs not to be a
zero temperature observable. In fact, the most useful quantity in our
calculation was the constant value of the free energy after the string
breaking/screening, at a fixed temperature. We kept this value 0 for all $a$,
that is we have chosen the constant $C(a)$ as
\begin{equation}
  2C(a) =  F_{\bar{q}q}(\r\to\infty,a,T_0),
\end{equation}
with a fixed $T_0$ (its value was $T_0=190$ MeV in the calculation).
The renormalized free energy therefore reads at any temperatures as
\begin{equation}
  F^\mathrm{ren}_{\bar{q}q}(\r,a,T) = F_{\bar{q}q}(\r,a,T) - F_{\bar{q}q}(\r\to\infty,a,T_0).
\end{equation}

\section{Results}

We used Symanzik improved gauge- and stout improved staggered fermionic actions.
The parameters of the action were the same as in Refs.~\cite{Aoki:2005vt}.
Table \ref{tab:lat} summarizes the lattices we used for the measurements.
These are the same gauge configurations as in Ref~\cite{Aoki:2006br}.
\begin{table}[htbp]
  \centering
  \begin{tabular}{|c|c|c|c|}
    \hline
    geometry & $\beta$ range & \# of $\beta$ values \\
    \hline
    \hline
    $16^3\times4$ & $3.2$ -- $3.425$ & 19\\
    \hline
    $24^3\times6$ & $3.45$ -- $3.705$ & 11 \\
    \hline
    $32^3\times8$ & $3.57$ -- $3.725$ & 7 \\
    \hline
    $48\times40^2\times 10$ & $3.63$ -- $3.86$ & 7 \\
    \hline
  \end{tabular}
  \caption{The lattices used for the Polyakov loop correlators.}
  \label{tab:lat}
\end{table}
We measured the Polyakov loop correlator for each possible $\r$ values which
could fit in the half-size of the spatial extent. Next we averaged the
correlator for distances $r=\sqrt{\r^2}$, including on- and off-axis
contributions. Note, that we take the continuum limit, where
rotation invariance should be restored. We binned the
data according to the lattice spacing, averaging Polyakov loop correlators
with the same $n = (int) (r+0.5)$. From the binned Polyakov loops we computed
the binned free energy as $F_{\bar{q}q,n} = \ln \exv{PP}_n/(N_t a)$. 
The $a(\beta)$
function was taken from the lines of constant physics determined earlier for
these set of lattices in \cite{Aoki:2006br,Aoki:2006we}. There the condition 
for the
determination of $\beta$ and the quark masses was to keep the ratios of the
physical values of $m_\pi,\, f_K$ and $m_K$ fixed at zero temperature. 

\begin{figure}[htbp]
  \centering
  \includegraphics[height=8.5cm,bb=17 170 587 615]{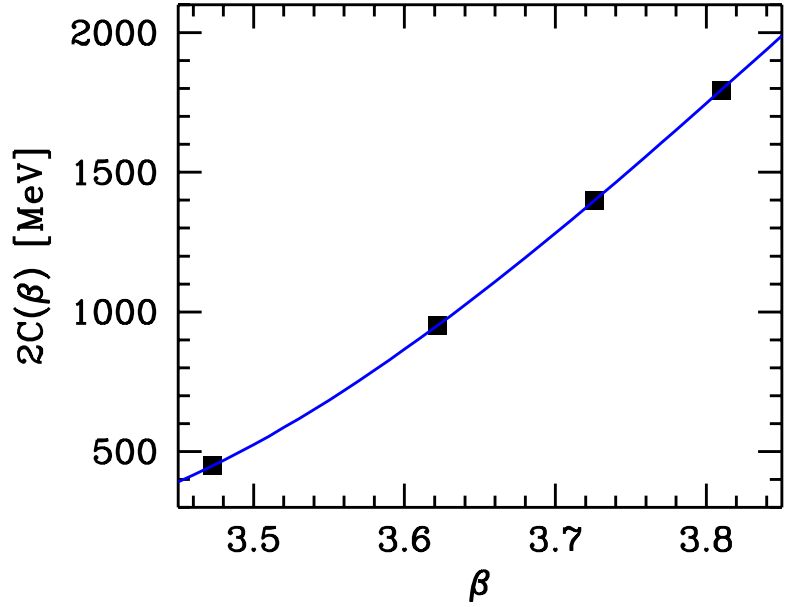}
  \caption{The additive renormalization factor to the free energy, as a function of $\beta$.}
  \label{fig:d_beta}
\end{figure}

The binned free energy was renormalized in the following way. For each $N_t$ and
$\beta$ we fitted the free energies with the function
\begin{equation}
  F_{fit}(r) = \frac{ae^{-br}}{r^c} +d.
\end{equation}
We then interpolated the fitted functions on each $N_t$ to the $\beta$
values corresponding to $T_0=190$~MeV. The asymptotic values of these
four ($N_t=4,6,8,10$) free energies gave $2C$
as a function of $\beta$. The four points and a fitted polynomial can be seen
on Fig. \ref{fig:d_beta}.
The value of $T_0=190$ MeV was
motivated by the fact that it lies already in the deconfined phase
where the statistical errors of the free energy are much smaller than
in the confined phase. At
this temperature the free energy at large distances, by definition, has no
lattice spacing dependence. At nearby temperatures we expect
similarly good behavior. 

\begin{figure}[htbp]
  \centering
  \includegraphics[height=8.5cm,bb=17 170 587 615]{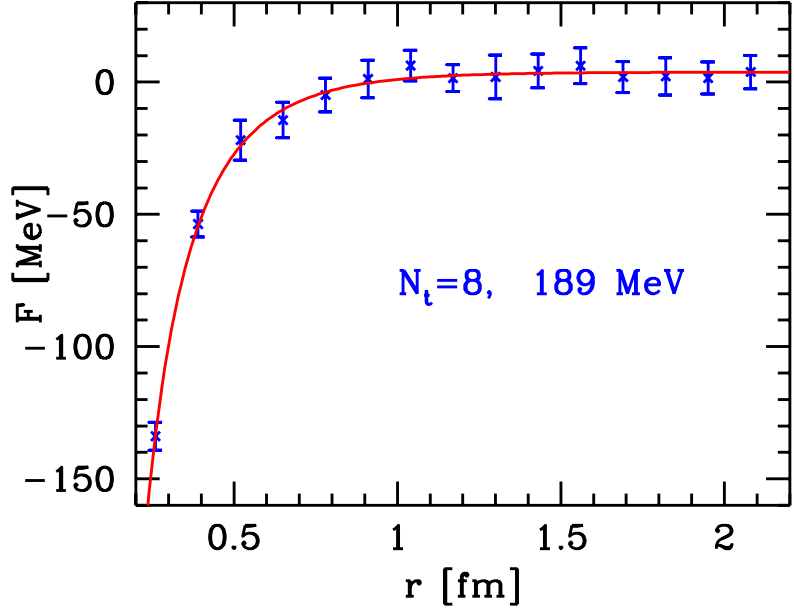}
  \caption{The renormalized free energy at $T=189$ MeV. The smooth curve is $F_{fit}$ discussed in the text.}
  \label{fig:renpot}
\end{figure}

Once we have the value of $2C(\beta)$ we can subtract it from all 
free energy values,
thus having their renormalized value. The result for temperature $T=189$ MeV
can be seen on Fig. \ref{fig:renpot} together with the fitted curve.
This $N_t=8$ point was the closest to the renormalization temperature $T_0$,
where we had raw data without interpolation.

\begin{figure}[htbp]
  \centering
  \includegraphics[height=8.1cm,bb=17 170 587 615]{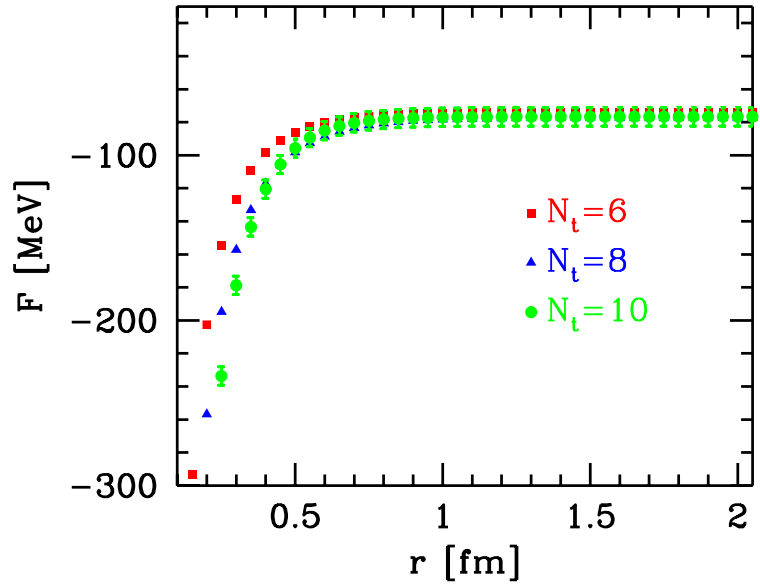}
  \caption{The renormalized free energies for $N_t=6,8$ and 10.}
  \label{fig:pot_nt}
\end{figure}

\begin{figure}[htbp]
  \centering
  \includegraphics[height=8.1cm,bb=17 170 587 615]{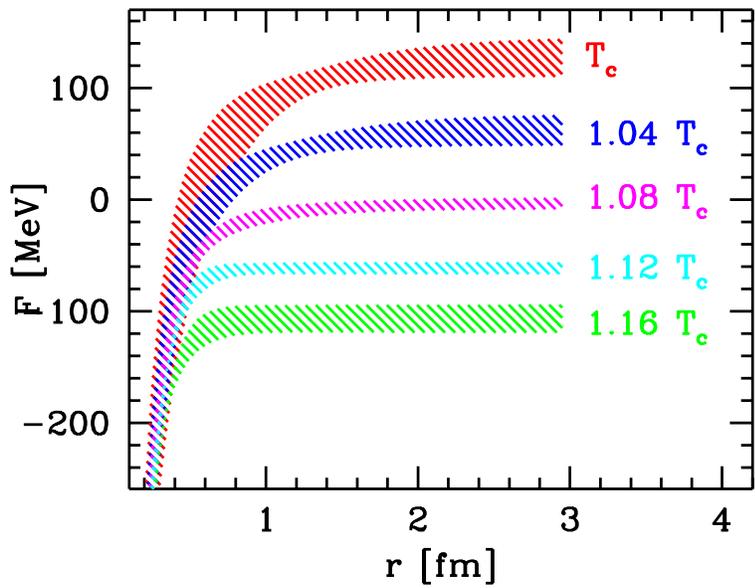}
  \caption{The renormalized free energies in the continuum limit.}
  \label{fig:pot}
\end{figure}

Since we now know the renormalized free energy for all lattice spacings, we 
can
take the continuum limit by using the $N_t=4,6,8$ and 10 free energies,
and extrapolate in $1/N_t^2 \sim a^2 \to 0$. In Fig. \ref{fig:pot_nt} one can
see the free energies at different $N_t$ values for $T=200$~MeV. We can see
that the lattice artefacts are small, $N_t=8$ and 10 results almost completely
coincide.
Therefore a safe extrapolation to $1/N_t^2=0$ is possible. We estimate the
systematic error of this extrapolation by comparing the results coming from
$N_t=6,8,10$ extrapolation and $N_t=8,10$ extrapolation. The result for the
renormalized free energy at different temperatures, including both the
statistical and the systematic errors, can be seen on Fig. \ref{fig:pot}.

\section{Conclusions}
We have determined the finite temperature renormalized static quark free
energy in QCD with dynamical staggered fermions using physical quark masses.
According to our expectations, the free energy is Coulomb-like for small
distances, at larger distances it is screened and/or exhibits string breaking,
and so flattens out. An important feature of the computation was the careful
renormalization procedure. We fixed a physical quantity: the asymptotic value
of the free energy at $T=190$ MeV, which was kept zero for all lattice
spacings. This defines the additive renormalization factor for the
quark-antiquark free energy as a function of the lattice spacing. At different
temperatures and different distances this factor must be used to renormalize
the free energy. The free energy defined in this way is monotonically
decreasing as a function of the temperature for all distances.

\section*{Acknowledgment}

Partial support of grants of \hbox{DFG F0 502/1}, \hbox{EU I3HP}, \hbox{OTKA
  AT049652} and \hbox{OTKA K68108} is acknowledged. A.J. is supported by the
Humboldt Foundation.

\end{document}